\documentclass[prl,twocolumn,twoside,showpacs,byrevtex,superscriptaddress,floatfix]{revtex4}

\usepackage{url}
\usepackage{graphicx,epsfig,verbatim,enumerate}
\usepackage{amssymb,amsmath}
\usepackage{ifthen}
\newboolean{twocolswitch}

\usepackage{color}

\usepackage{rotating}
\usepackage{longtable}
\usepackage{array}

\newcommand{\sindex}[1]{}
\newcommand{\nindex}[1]{}

\newcommand{\etal}{\textit{et al.}}
\newcommand{\www}[1]{\url{#1}}

\newcommand{\Req}[1]{Eq.~(\ref{#1})}

\newcommand{\yes}{}
\newcommand{\no}{}

\newcommand{\dee}[1]{\mbox{d}#1}

\newcommand{\PreserveBackslash}[1]{\let\temp=\\#1\let\\=\temp}
\newcommand{\PBS}[1]{\let\temp=\\#1\let\\=\temp}

\newcommand{\volume}[1]{\Omega_{d,D}(#1)}

\newcommand{\Vnet}{V_{\rm net}}

\newcommand{\gammamax}{\gamma_{\rm max}}

\newcommand{\sinkdensity}{\rho}

\setboolean{twocolswitch}{true}

\begin{document}

\title{
  On the optimal form of branching supply and collection networks

}

\author{
\firstname{Peter Sheridan}
\surname{Dodds}
}
\email{peter.dodds@uvm.edu}

\affiliation{
  Department of Mathematics and Statistics,
  Center for Complex Systems,
  \&
  the Vermont Advanced Computing Center,
  University of Vermont,
  Burlington,
  VT, 05401
}

\date{\today}

\begin{abstract}

For the problem of efficiently supplying material to 
a spatial region from a single source,
we present a simple scaling argument 
based on branching network volume minimization
that identifies limits to the scaling
of sink density.
We discuss implications for two fundamental
and unresolved problems in organismal biology and geomorphology:
how basal metabolism scales with body size for homeotherms
and the scaling of drainage basin shape on eroding landscapes.

\end{abstract}

\pacs{89.75.Fb,89.75.Hc,87.19.U-,92.40.Gc}

\maketitle

In both natural and man-made systems,
branching networks universally facilitate
the essential task of supplying material
from a central source to a widely distributed 
sink population.
Branching networks also
underlie the complementary process
of collecting material from many sources
at a single sink.
Such networks typically exhibit structural 
self-similarity over many 
orders of magnitude:
river networks drain continents~\citep{horton1945a,tokunaga1966a,rodriguez-iturbe1997a},
arterial and venal networks move blood between the macroscopic heart
and microscopic capillaries~\citep{fung1990a},
and trees and plants orient leaves in space
taking on the roles of both structure and transportation.

We address the following questions regarding supply networks:
\textbf{(1)} What is the minimum network volume required
to  continually supply material from a source to a population of 
sinks in some spatial region $\Omega$?;
and
\textbf{(2)} How does this optimal solution scale if
$\Omega$ is rescaled allometrically?
(For convenience, we use the language of 
distribution, i.e., a single source supplying many sinks.)
Our approach is inspired by that of Banavar~\etal~\citep{banavar1999a,banavar2002a}
who sought to derive scaling properties of optimal transportation
networks in isometrically growing regions based on a flow rate argument;
Banavar~\etal's approach followed the seminal
work of West~\etal~\citep{west1997a}
who suggested supply networks were
key to understanding the metabolic limitations of organisms,
and focused on network impedance minimization
(see~\citep{dodds2001d,makarieva2003a}).
In contrast to this previous work,
our treatment is explicitly geometric.
We also accommodate four other key features:
the ambient dimension, allometrically growing regions, variable sink density,
and varying speed of material transportation.

\begin{figure}[htbp]
  \centering
  \includegraphics[width=0.5\textwidth]{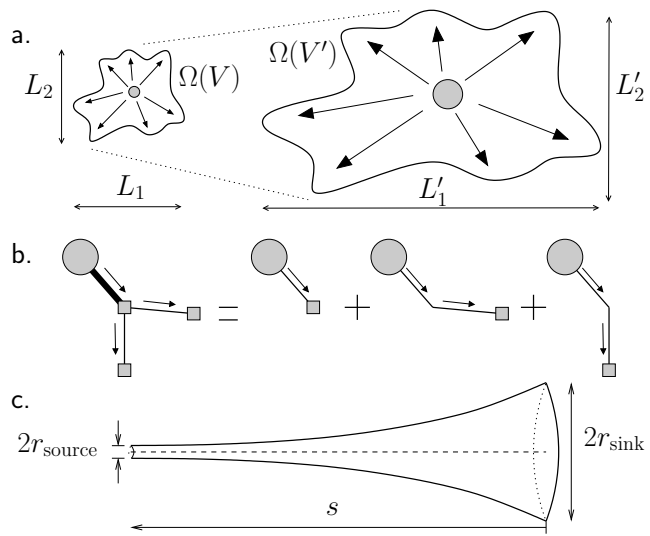}
  \caption{
    \textbf{(a)}
        We consider families
    of $d$-dimensional spatial regions that scale allometrically 
    with $L_i \propto V^{\gamma_i}$, and
    exist in a $D$-dimensional space
    where $D \ge d$.
    For the $d$=$D$=2 example shown, $\gammamax = \gamma_1 > \gamma_2$,
    and $L_1$ grows faster than $L_2$.
    We require that each spatial region is star-convex,
    i.e., from at least one point 
    all other points are directly observable, 
    and the single source must be located at any one
    of these central points.
    \textbf{(b)}
    Distribution (or collection) networks can be thought of as 
    a superposition of virtual vessels.  In the example shown,
    the source (circle) supplies material to the three sinks (squares).
    \textbf{(c)}
    Allowing virtual vessels to expand as they move away from the
    source captures a potential decrease in speed in material flow.
    For scaling of branching network form to be affected, the 
    radius $r$ of a virtual vessel must
    scale with vessel length $s$ (measured
    from the sink) as $s^{-\epsilon}$.
  }
  \label{fig:supnet.shape}
\end{figure}

We consider the problem of network supply for
a general class of $d$-dimensional 
spatial regions in a $D \ge d$ dimensional space.
Each region $\Omega$ has volume $V$ and
overall dimensions $L_1\times L_2 \times \cdots \times L_d$
(see Fig.~\ref{fig:supnet.shape}a).
We allow these length scales to 
scale as $L_i \propto V^{\gamma_i}$, creating families of allometrically
similar regions.
For isometric growth, all dimensions scale uniformly
meaning $\gamma_i = 1/d$, 
while for allometric growth, 
we must have at least one of the $\{\gamma_i\}$ being different
(Fig.~\ref{fig:supnet.shape}b).
For the general case of allometry, we choose an ordering of 
$\{\gamma_i\}$ such that the length scales are arranged from
most dominant to least dominant: 
$\gammamax = \gamma_1 \ge \ldots \ge \gamma_d$.

We assume that isolated sinks are located
throughout a contiguous spatial region $\Omega$ (volume $V$)
which contains a single source located at $\vec{x} = \vec{0}$.
We allow sink density to follow
$\sinkdensity \sim \sinkdensity_0(V)(1+ a ||\vec{x} ||)^{-\zeta}$
where $a$ is fixed, $\zeta \ge 0$, and $||\vec{x}||$ is 
the distance from the source.
When the exponent $\zeta=0$, $\sinkdensity$ is constant
throughout the region (as for capillaries in organisms),
but remains a function of the region's overall volume $V$.
While decreasing sink density ($\zeta>0$)
does not reflect the reality of biological organisms, 
it is not an unreasonable postulate
for other supply/collection systems, and may be of use
in modelling transportation to and from cities.
Last, we assume each sink draws approximately
the same amount of material from the source per unit time.
The material travelling from the source
to a specific sink takes up a certain volume
of the network, 
and while this volume of material may not
be coherent away from the sink, 
we can nevertheless imagine 
separate `virtual vessels' transporting material
from the source to the sinks (see Fig.~\ref{fig:supnet.shape}b).
(Only in the smallest, outer branches will virtual vessels
coincide with physical vessels.)
Material flow rate will then vary according to changes in
the cross-sections of these vessels.

We take the cross-sectional area
of these virtual vessels to be bounded
by a fixed upper limit at the sink
(see Fig.~\ref{fig:supnet.shape}b).
We allow that material speed may
increase with proximity to the source,
meaning these virtual vessels may taper.
If the radius decreases as $r_{\rm sink}(1+c s)^{-\epsilon}$,
where $s$ is the length of the vessel as
measured from the sink and $c$ is a constant, then the volume
of a virtual vessel grows as $v_{\rm vessel} \sim s^{1-2\epsilon}$ for 
$0 \le \epsilon < 1/2$ and $v_{\rm vessel} \sim s^{0}$ for $\epsilon \ge 1/2$
(we can therefore focus on $\epsilon = 1/2$ to represent the latter case).
We ignore all other possible taperings since only an
algebraic decay relationship between vessel radius and length
will affect the scaling of overall network volume $\Vnet$.
If however, there is a minimum virtual vessel radius
(i.e., a limit to material speed) then vessel volume
must grow linearly with length: $v_{\rm vessel} \sim s^{1}$.

The overall network volume $\Vnet$ is 
the sum of all virtual vessel volumes,
and is evidently minimized when virtual vessels
travel directly from the source to each sink---the extreme case of a star network.
While real, large-scale distribution networks are 
branched, many examples are close to this limit
in terms of path length~\citep{gastner2006a}.
Minimal network volume therefore grows as
\begin{equation}
  \label{eq:supnet.optVn1}
  \min \Vnet  \propto 
  \int_{\volume{V}} 
  \sinkdensity_0
  (1+a || \vec{x} ||)^{-\zeta}
  \, 
  ||\vec{x}||^{1-2\epsilon}
  \,
  \dee \vec{x},
\end{equation}
where $0 \le \epsilon \le 1/2$,
and we have indicated a spatial region 
$\Omega$ scaled to have volume $V$ by $\volume{V}$.
The integral's leading order behavior gives
the optimal scaling of $\Vnet$ with $V$:
\begin{equation}
  \label{eq:supnet.optVn4}
  \min \Vnet
  \sim
  \sinkdensity_0 V^{1+\gammamax(1-2\epsilon-\zeta)},
\end{equation}
where again $\sinkdensity_0 = \sinkdensity_0(V)$.
When $\epsilon \ge 1/2$, 
$\min \Vnet \sim \sinkdensity_0 V^{1-\zeta\gammamax}$.
The scaling of minimal network volume with $V$ is thus governed by 
sink density $\sinkdensity$, vessel scaling,
and the dominant length scale through $\gammamax$,
and we first address the role of the latter.
Since for isometric scaling, $\gammamax = 1/d$, 
whereas for allometric scaling, $\gammamax > 1/d$,
we immediately see that from a scaling perspective,
isometrically growing regions require less network volume 
than allometrically growing ones, and are in this sense
more efficiently supplied.
Efficiency also increases with the dimension $d$
since network volume scales more closely with overall volume 
($\gammamax = 1/d$ decreases).
Furthermore,
shapes that scale allometrically effectively function as lower-dimensional,
isometrically scaling objects and are therefore less efficiently supplied
(the equivalent spatial dimension is $1/\gammamax$).

We see from \Req{eq:supnet.optVn4} that
network volume straightforwardly 
increases linearly with $\sinkdensity_0$.
How $\sinkdensity_0$ in turn scales with $V$ depends 
on the specific system, and in particular on whether $D=d$ or $D>d$.
We now specialize our general result
for the two cases of blood networks and river networks.

\textbf{Blood networks [$D$=$d$=3]:}
If material is costly, as in the case of blood, then we expect
that isometric scaling ($\gammamax = 1/d$) to be
attained by evolution.
We take $\zeta=0$ because capillaries (the sinks) are distributed
relatively uniformly.  Furthermore, since both blood
velocity in the aorta and in capillaries
change little with $V$ during resting states~\citep{west1997a,weinberg2006a}, 
we have $\epsilon=0$.
Lastly, it is well observed that the volume of 
blood scales linearly with organismal volume~\citep{stahl1967a},
$\Vnet \propto V$.
(In general, for $D=d$,
we must have $\Vnet \propto V$ as otherwise we would have
the nonsensical
limits of $\Vnet/V \rightarrow 0$ or $\infty$ as $V \rightarrow \infty$.)
Since we already have that $\Vnet \propto \sinkdensity_0 V^{1+1/d}$,
the additional constraint $\Vnet \propto V$
means sink density must decrease as volume increases:
$\sinkdensity_0  \propto V^{-1/d}$, where
$\sinkdensity_0$ for resting organisms now
refers to the effective or active sink density
(organisms at rest have a substantial proportion of inactive
capillaries that are called into use during higher rates of overall
activity~\citep{hoppeler1981a}).

It follows that $P_{\rm rest}$, the average rate of energy 
use in a resting state (basal power), which is proportional to the
number of active sinks in $\Omega$, can at best scale 
as $P_{\rm rest} = \sinkdensity V \propto V^{-1/d} V \propto M^{(d-1)/d}$,
where we have assumed that $V$ scales as body mass $M$.
For three dimensional organisms, we therefore have 
\begin{equation}
  \label{eq:supnet.metabscaling}
  P_{\rm rest} \propto M^{2/3}.
\end{equation}
If organism shapes obey instead an allometric scaling 
then power scales more slowly as 
$P_{\rm rest} \propto M^{1-\gammamax}$ with $\gammamax > 1/3$,
contrary to the McMahon's theory of elastic similarity~\citep{mcmahon1973a}.

We note that in detail, blood networks do not appear to have
universal forms~\citep{huang1996a}, showing substantial
variation in branching structure across and within species; 
we therefore argue that it is only the system level that matters
and that branching networks need only approximate star networks.

Crucially, the scaling law of \Req{eq:supnet.metabscaling} balances
with the standard one based on organismal surface area $S$.
For homeothermic organisms, who must constantly balance heat loss
to maintain a steady internal temperature, we have that $P_{\rm rest} \propto S$
(due primarily to radiation but also convection~\citep{hardy1937a}).  
For isometrically scaling organisms, $S \propto M^{2/3}$,
and this is well supported empirically~\citep{stahl1967a}.
Moreover, it is easy to show that only isometrically growing
shapes balance $P_{\rm rest}$ since the $d-1$ dimensional surface area
of a growing region $\Omega$ scales as $V^{1-\gamma_{\min}}$,
and therefore $\gammamax = \gamma_{\min}=1/d$.
Thus, the most efficient network in terms of minimal volume
is also the one that precisely balances radiative heat loss.

Our seemingly reasonable result, which was empirically
observed over a century ago by Rubner~\citep{rubner1883a},
runs counter to nearly eighty years of reports that $P_{\rm rest} \propto M^{3/4}$.
Kleiber~\citep{kleiber1932a} first suggested the exponent might be 3/4
in the 1930's after measuring a value of 0.76 for 13 mammals
(his practical reason for choosing 3/4 was 
to simplify slide rule calculations~\citep{schmidt-nielsen1984a}).
In the decades following, a general but not universal consensus
on a  ``3/4-law of metabolism'' was reached~\citep{blaxter1965a,west1997a,lane2005a}
The issue remains controversial both theoretically~\citep{dodds2001d,savage2008a,kozlowski2004a,brown2005a},
and empirically:
some recent statistical analyses have shown that a 2/3 exponent is
well supported by large data sets 
for warm-blooded organisms (both birds and mammals)~\citep{heusner1991a,dodds2001d,white2003a,white2005a}
while others have found evidence in favor of a 3/4 exponent or
no simple scaling relationship~\citep{savage2004a,dodds2001d,white2006a,white2007a,packard2008a};
and for cold-blooded organisms, plants, and invertebrates, 
a much broader range of exponent values have been measured empirically and
predicted from theory~\citep{patterson1992a,glazier2005a,reich2006a,glazier2006a}.

Perhaps the most important aspect of the 2/3 vs.\  3/4 debate
is that from an optimization point of view, 
the lower the scaling of resting metabolism the better.
A scaling of $M^{3/4}$ (or any power exceeding 2/3)
would point to either a fundamental scaling limitation
for warm-blooded organisms, or to the existence of 
a cost other than volume minimization,
such as impedance~\citep{west1997a}.
Where a higher exponent would be desirable is in
the scaling of maximal power $P_{\rm max}$,
which is unsustainable and depends on stored energy, and indeed,
$P_{\rm max}$ scales almost linearly with mass~\citep{glazier2005a}.

\begin{figure}[tbhp]
  \centering
  \includegraphics[width=0.5\textwidth]{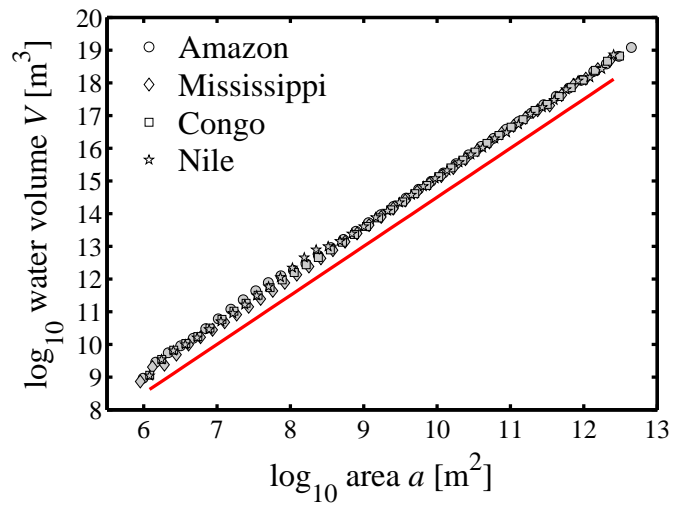}
  \caption{
    The scaling of network volume versus basin area 
    for four continental-scale river networks
    The solid line indicates a scaling of 3/2.
    Network volume is estimated for
    an idealized steady-state, uniform rainfall condition
    normalized such that $\Vnet = \sum_{ij} a_{ij}$, where $a_{ij}$ 
    is the area of the basin draining into the $ij$th cell 
    on a coarse-grained version of a landscape,
    and the data is binned in log space~\citep{banavar1999a}.
  }
  \label{fig:supnet.watervolume02}
\end{figure}

\textbf{River networks [$D$=3 $>$ $d$=2]:}
The patterns of large-scale river networks have long
drawn scientific interest, naturally from hydrologists and 
geomorphologists~\citep{horton1945a,rodriguez-iturbe1997a},
but also from statistical physicists seeking evidence
of universality in nature~\citep{maritan1996a,dodds1999a,banavar2001a,maritan2002a}.
In our framework, river networks are collection systems:
water flows from many sources (channel heads~\citep{montgomery1992a}) 
to a single sink, the outlet of the network's main stream.  
The description of river 
network geometry has often focused on Hack's law~\citep{hack1957a}
which relates the area $a$ of a drainage basin to the length $l$ of 
its longest stream: $l \propto a^{h}$.
Various studies of small-scale basins, starting with 
Hack's initial work~\citep{hack1957a,rodriguez-iturbe1997a},
have suggested that the `Hack exponent' $h$ exceeds 1/2, 
indicating an anomalous allometric scaling of basin shape wherein large
basins are relatively long and thin compared to smaller ones.
Indeed, for sufficiently small, homogeneous landscapes, there may be
a universal value of $h > 1/2$, yet to be fully understood
theoretically~\cite{maritan1996a,maritan2002a,banavar2001a}.
However, the most comprehensive data sets robustly show that $h=1/2$
for large-scale networks;  in particular,
Montgomery and Dietrich~\citep{montgomery1992a}
found that over twelve orders of magnitude variation
in basin area, $l \propto a^{0.49}$ (their data set
mixed both Euclidean overall basin length $L$ and main
stream length $l$; generally, $l \propto L^{\beta}$ with
$\beta$ very close to, if not equal to, unity).

The empirical observation that $h=1/2$ accords with
our result that with respect to network volume minimization, 
isometrically growing regions are most efficient;
with our optimality argument, this becomes a stronger statement than 
appealing only to dimensional analysis.  

Beyond isometry, we have the scaling of network
volume to consider.  We now take $\sinkdensity_0$ to
be constant and again set $\zeta=0$, meaning we assume that,
when averaged over time, rain falls approximately
uniformly across a landscape.  We also assume 
$\epsilon=0$ for the continental scale networks
we examine below.  In contrast to the case of cardiovascular networks,
the constraint that network volume must scale as overall
volume (or basin area $a$) cannot apply to river networks,
since 
\Req{eq:supnet.optVn4} now gives 
$\Vnet \propto \sinkdensity_0 V^{1+1/d} \propto V^{3/2} = a^{3/2}$.
The reason is simple: river networks lie on a $d=2$ dimensional
surface embedded in $D=3$ dimensions, and the presence
of a third dimension allows the total water in the network 
to grow faster than if the embedding
dimension was $D=2$.

In Fig.~\ref{fig:supnet.watervolume02} we
show that $\Vnet$ scales as $a^{3/2}$ 
for four continental-scale networks: the Mississippi, the Amazon,
the Congo, and the Nile~\citep{dodds2001a}.
The scaling is robust and holds over
6 and 10 orders of magnitude in area and network volume
respectively, indicating that self-similar drainage basins 
most efficiently drain large-scale landscapes.

While we have argued that the optimal scaling corresponds
to the isometric case of $h=1/2$, we can directly see this connection arising
from known scaling laws of river networks~\citep{maritan1996a,dodds1999a}.
As put forward in~\cite{maritan2002a},
network volume scales with basin area as
$\Vnet \propto a^{1+h}$, showing that for the optimal 
case, we indeed have $h=1/2$.
Note that only by considering allometrically growing regions,
as we do here, can this connection be formally made.

We acknowledge that a stronger optimization may be at work for eroding landscapes.
In particular, previous theoretical analyses
suggest that in terms of energy minimization, landscapes may
reach local, dynamically accessible minima~\cite{banavar2001a,maritan2002a},
though sufficient self-averaging 
inherent in larger landscapes may render such minima
indistinguishable from the global one.
Nevertheless, our argument and data analysis show that for large-scale networks
on heterogeneous landscapes, far exceeding the typical correlation
length for precipitation events, a kind of optimal volume minimization 
is achieved.

With suitable modifications, 
our findings may be found relevant to other systems, in particular to plants 
(when seen as two connected branching networks),
as well as to the scaling limits of episodic
movement such as the transportation of people in and out of city centers.
(Assuming constant flow as we do here, then a decay of
sink density with $\zeta=1$ follows for growing cities
where transportation remains unchanged with city size.)
While we have demonstrated that empirical evidence supports
a geometric optimality for two kinds of large-scale natural branching
networks, not all systems will be optimal or may be optimized.
For example, if allometric scaling of organismal shape
is demanded by some other 
constraints (e.g., due to the effects of gravity), 
then  blood volume will be forced to
obey a poorer scaling with overall volume~\citep{makarieva2003a}.

\end{document}